\documentclass[
twocolumn,
amsmath, amssymb, amsfonts,preprintnumbers,aps,prd,longbibliography,nofootinbib]{revtex4-2}

\usepackage{graphicx}
\pdfoutput=1

\usepackage{dcolumn}
\usepackage{bm}
 \usepackage{amstext}
 \usepackage{amssymb}
 \usepackage{amsmath}

 \usepackage{color}
 \usepackage{bbold}
 \usepackage{delimset} 
\usepackage[caption=false,justification=justified]{subfig}
\usepackage[colorlinks=true]{hyperref}
\usepackage{orcidlink}
\usepackage{float}
\usepackage[normalem]{ulem}
\usepackage{mathtools}
\usepackage[colorlinks=true]{hyperref}

\usepackage{tikz}
\usetikzlibrary{decorations.pathmorphing}
\usetikzlibrary{decorations.markings}
\usetikzlibrary{positioning, shapes, snakes}
\usetikzlibrary{arrows.meta}

\begin{document}

\author{Yong Zhang \!\orcidlink{0000-0002-3522-0885}}

\email{zhangyong1@nbu.edu.cn}

\affiliation{
Institute of Fundamental Physics and Quantum Technology\\ \& School of Physical Science and Technology, Ningbo University, Ningbo, Zhejiang 315211, China }

\title{Local asymmetry in interference as a probe of quantum probability}

\begin{abstract}
Quantum interference provides one of the most sensitive probes of quantum mechanics. While linear superposition fixes the positions and quadratic curvature of interference fringes, it remains unclear whether the probabilistic postulate itself—the Born rule—can be tested through finer, local features of interference patterns. Here we show that a minimal deformation of quantum probability gives rise to a robust and symmetry-protected signature: a left--right asymmetry in the local shape of interference fringes.  Remarkably, this effect leaves the linear Schr\"odinger dynamics intact and does not shift fringe positions or modify their quadratic curvature.  Instead, it appears exclusively as a cubic skewness of local intensity profiles, providing a clean and falsifiable observable. We demonstrate this behavior within a controlled realization that preserves linear dynamics while minimally deforming the probabilistic assignment. The resulting signature is universal, scale insensitive, and cannot be mimicked by conventional sources of experimental noise.  Our results identify local asymmetry in interference as a direct probe of quantum probability itself, suggesting that features often regarded as removable imperfections may encode fundamental information beyond fringe positions and widths.
\end{abstract}

\maketitle 
\section*{Introduction}
Quantum interference lies at the heart of quantum mechanics.
From Young’s double-slit experiment to modern interferometric platforms,
the superposition principle and linear time evolution give rise to
interference fringes whose positions and overall shapes are among the most
precisely tested predictions of the theory.
In standard quantum mechanics, these phenomena are interpreted through the
quadratic Born rule, which assigns probabilities as the modulus squared of
a complex wave function~\cite{SakuraiQM}.

Despite its central role, the Born rule itself is rarely subjected to direct
experimental scrutiny.
Most precision tests of quantum mechanics focus on the linear dynamics or on
symmetry principles, while the probabilistic postulate is typically taken
as fixed.
Interference experiments, in particular, are almost exclusively analyzed in
terms of fringe positions, visibility, or quadratic broadening, implicitly
assuming the exact validity of the Born rule at all levels of detail.
Finer, local features of interference patterns are usually treated as
removable imperfections rather than as potential physical observables.

A natural question therefore arises: can the probabilistic structure of
quantum mechanics be probed directly through such local features of
interference patterns?

In this work we show that it can.
We demonstrate that a minimal deformation of quantum probability leads to a
distinct and robust interferometric signature: a protected left--right
asymmetry in the local shape of bright fringes.
Remarkably, this effect leaves the interference conditions unchanged.
It does not shift fringe positions, nor does it modify their quadratic
curvature.
Instead, it manifests itself exclusively through a cubic skewness of the
local intensity profile, providing a clean observable that is sharply
separated from conventional sources of experimental noise.

At a formal level, the deformation we consider can be stated directly at the
level of the probabilistic assignment.
Instead of the standard quadratic Born rule $P = |\psi|^2$, we allow the
minimal generalization
\begin{equation}
P(\psi)
=
\psi^{\star\,(1-i\theta)}\,
\psi^{\,1+i\theta},
\end{equation}
which reduces continuously to $|\psi|^2$ as $\theta \to 0$.
This deformation preserves positivity and admits a smooth classical limit.

One convenient realization of such a deformation, which we use for
definiteness, is to allow the phase--action scale $\kappa$ to acquire a
parametrically small imaginary component, such that
\begin{equation}
\theta \equiv \frac{\mathrm{Im}\,\kappa}{\mathrm{Re}\,\kappa}.
\end{equation}
In this sense, the quantity playing the role of Planck’s constant in the
phase evolution is allowed to be slightly complex, while its real part
remains fixed.
Crucially, this modification leaves the linear structure of quantum dynamics
unchanged: wave functions continue to superpose linearly, canonical
operators remain well defined, and probability conservation follows from
the continuity equation.

From an experimental perspective, this separation is essential.
The deformation does not alter interference conditions, fringe positions,
or quadratic widths.
Instead, its effects appear only at the level of local probability
distributions, where they induce subtle asymmetries that vanish smoothly in
the limit $\theta \to 0$.
Interference experiments therefore provide a natural and controlled setting
in which to probe possible deviations from the standard Born rule, without
invoking new dynamical assumptions.

While the linear dynamics remains intact, the probabilistic assignment is
modified in a controlled and universal manner.
The effects governed by $\theta$ are scale insensitive and enter exclusively
through phases.
As a result, interference phenomena provide a uniquely sensitive arena for
detecting such deviations.
We show that the deformation preserves the locations and quadratic curvature
of interference fringes, while inducing a characteristic local asymmetry
that cannot be mimicked by phase noise, path-length fluctuations, or
finite detector resolution.
This establishes local fringe skewness as a direct probe of quantum
probability itself, rather than of the underlying linear dynamics.

\section*{Schr\"odinger dynamics with complex $\kappa$ and a $\theta$-deformed Born rule}

Allowing the phase--action scale $\kappa$ to take complex values leads to a
straightforward extension of the Schr\"odinger equation,
\begin{equation}
i\kappa\,\partial_t \psi
=
-\frac{\kappa^2}{2m}\,\nabla^2 \psi
+ V\,\psi ,
\label{eq:sch_complex_kappa}
\end{equation}
which remains linear in the wave function $\psi$ for arbitrary complex
$\kappa$.
This modification introduces a single new dimensionless parameter,
$
\theta \equiv {\mathrm{Im}\,\kappa}/{\mathrm{Re}\,\kappa},
$
while leaving the linear structure of quantum dynamics unchanged.
In particular, the superposition principle continues to hold, and standard
operator-based spectral constructions remain applicable.

Equation~\eqref{eq:sch_complex_kappa} admits a transparent phase--amplitude
representation,
\begin{equation}
\psi = R\,e^{iS/\kappa},
\end{equation}
in terms of which the dynamics can be rewritten as two coupled equations for the real fields $R$ and $S$:
a modified Hamilton--Jacobi equation,
\begin{equation}
\partial_t S
+ \frac{(\nabla S)^2}{2m}
+ V
+ Q[R]
= 0 ,
\qquad
Q[R]
\equiv
-\,\frac{\kappa^2}{2m}\,\frac{\nabla^2 R}{R} ,
\label{eq:modified_HJ}
\end{equation}
together with the continuity equation,
\begin{equation}
\partial_t (R^2)
+ \nabla\!\cdot\!\left(
R^2\,\frac{\nabla S}{m}
\right)
= 0 .
\label{eq:continuity}
\end{equation}
These equations remain first order in time and are well defined for complex
$\kappa$.
Importantly, probability conservation follows directly from the continuity
equation~\eqref{eq:continuity}, independent of the value of $\theta$.

From the perspective of linear quantum mechanics, canonical operators are
defined exactly as usual.
For example, the momentum field in the phase--amplitude representation is
$p=\nabla S$, which translates into the linear operator
\begin{equation}
\hat p = -i\kappa\,\nabla .
\end{equation}
This operator remains well defined for complex $\kappa$ and generates real
momentum eigenvalues.
The same reasoning applies to orbital angular momentum,
\begin{equation}
\hat L = -i\kappa\,\mathbf{r}\times\nabla ,
\end{equation}
whose algebra follows from spatial symmetry alone.

Time evolution is generated by the linear Hamiltonian operator
\begin{equation}
\hat H
=
-\frac{\kappa^2}{2m}\,\nabla^2 + V ,
\end{equation}
which is generally non-Hermitian for complex $\kappa$.
Energy eigenvalue problems therefore retain their standard form,
\begin{equation}
\hat H \psi_n = E_n \psi_n ,
\end{equation}
with time dependence
\begin{equation}
\psi_n(\mathbf{x},t)
=
\psi_n(\mathbf{x})\,\exp\!\left(-\,\frac{iE_n t}{\kappa}\right).
\end{equation}
The real part of $E_n$ controls oscillatory phase evolution, while the
imaginary part sets an intrinsic linewidth.
Thus, the familiar spectral interpretation of quantum mechanics remains
intact.

While the linear dynamics is unchanged, the probabilistic assignment is
modified.
The probability density is defined as $R^2$, as dictated by the continuity
equation~\eqref{eq:continuity}, and when expressed in terms of the complex
wave function it takes the form
\begin{equation}
P(\psi)
\equiv R^2
=
\psi^{\star\,(1-i\theta)}\,
\psi^{\,1+i\theta}
.
\end{equation}
This $\theta$-deformed Born rule reduces continuously to the standard
quadratic form as $\theta\to0$, while preserving positivity and probability
conservation.
In this way, a minimal modification of the phase scale naturally induces a
controlled deformation of quantum probability, without altering the linear
structure of the theory.

\section*{Interference and a protected local asymmetry}

Allowing the phase--action scale $\kappa$ to acquire a small imaginary
component leaves the linear dynamical structure of quantum mechanics intact,
while modifying its probabilistic interpretation.
In particular, the standard quadratic Born rule no longer applies, and
probability assignments become explicitly sensitive to the phase structure of
the wave function.

This feature is most transparently revealed in interference phenomena.
As a minimal setting, consider a superposition of two wave packets,
\begin{equation}
\label{two_wave_packets}
\psi = \psi_1 + \psi_2,
\qquad
\psi_j = R_j\,e^{iS_j/\kappa}.
\end{equation}
In the limit $\theta\to0$, the probability distribution reduces to the familiar
two-path interference pattern,
\begin{equation}
\label{P0}
P_0
=
R_1^2 + R_2^2
+
2R_1R_2\cos\frac{S_1-S_2}{\mathrm{Re}\,\kappa},
\end{equation}
with interference fringes located at fixed values of the relative phase
$S_1-S_2$.

For finite $\theta$, the probabilistic assignment is modified to
\begin{equation}
\label{theta_bornrule_arg}
P(\psi)
=
\psi^{\star\,(1-i\theta)}\,
\psi^{\,1+i\theta}
=
|\psi|^2\,e^{-2\theta\,\arg\psi},
\end{equation}
which follows directly from the continuity equation and reduces smoothly to
$P_0$ as $\theta\to0$.
Importantly, the deformation depends only on the local phase $\arg\psi$ and
does not alter the linear superposition principle or probability conservation.

Two immediate structural consequences follow.
First, the locations of the interference fringes remain unchanged:
maxima are still fixed by the condition
\begin{equation}
S_1-S_2 = 2n\pi\,\mathrm{Re}\,\kappa .
\end{equation}
Second, the quadratic curvature of each bright fringe is identical to that of
standard quantum mechanics.
The deformation therefore does not induce any rigid fringe shift or symmetric
broadening.

The leading observable effect instead appears in the local shape of the bright
fringes.
Expanding around a maximum,
\begin{equation}
\frac{S_1-S_2}{\mathrm{Re}\,\kappa} = 2 n\pi + \sigma,
\qquad |\sigma|\ll1 ,
\end{equation}
one finds that the undeformed probability distribution is purely quadratic in
$\sigma$, while the leading $\theta$-dependent correction is cubic in the local phase deviation and odd under $\sigma\to-\sigma$,
\begin{equation}
\label{deltaPsim}
\delta P 
\sim
\frac{2}{3}
\theta R_1R_2 \frac{R_1-R_2 }{R_1+R_2} \sigma^3
.
\end{equation}
As a result, the bright fringe acquires a left--right asymmetric distortion,
while its position and width remain unchanged.

This skewness requires both a nonvanishing deformation parameter $\theta$ and
an amplitude imbalance between the interfering components.
In the perfectly symmetric case $R_1=R_2$, the effect vanishes identically.
Conversely, for generic interference configurations with spatially varying
amplitudes, the cubic asymmetry represents the dominant $\theta$-dependent
signature.
Because it affects neither fringe locations nor quadratic curvature, the
skewness cannot be mimicked by conventional sources of phase noise,
path-length fluctuations, or finite detector resolution, which typically
produce symmetric distortions.

Finally, the structural origin of this effect implies that it is not restricted
to the superposition of two wave packets.
The absence of any $\theta$-dependent shift of interference maxima and of any
quadratic modification of fringe curvature follows from the fact that the
deformation enters exclusively through the phase $\arg\psi$, rather than
through the modulus $|\psi|$.
As a result, in more general superpositions involving multiple interfering
components, the dominant interference peaks remain fixed in position and width,
while $\theta$-dependent effects appear only through local, higher-order
distortions of their intensity profiles.
In this sense, the cubic skewness of bright fringes constitutes a generic and
structurally stable probe of a $\theta$-deformed quantum probability rule.

A detailed derivation of \eqref{deltaPsim} is provided in the Supplementary
Information.

\section*{An experimentally accessible observable}

The interferometric consequence of a $\theta$-deformed Born rule identified
above admits a particularly simple and robust experimental characterization.
Because the deformation leaves both the locations and the quadratic curvature
of interference fringes unchanged, conventional observables such as fringe
spacing, visibility, or linewidth are insensitive to $\theta$ at leading
order.
The relevant information is instead encoded in the \emph{local asymmetry} of
individual bright fringes.

A natural observable is therefore provided by a normalized measure of
left--right skewness of a fringe intensity profile.
Concretely, consider a bright fringe centered at a position $x_0$ and define
the local deviation coordinate $\delta x = x-x_0$.
Any symmetric distortion of the fringe contributes only to even powers of
$\delta x$, while the leading $\theta$-dependent effect enters as an odd,
cubic contribution.
As a result, quantities that isolate the odd moments of the local intensity
distribution are directly sensitive to $\theta$.

One convenient choice is the normalized third central moment of the intensity
profile,
\begin{equation}
\mathcal{S}
\equiv
\frac{
\int d(\delta x)\, (\delta x)^3\, P(x)
}{
\left[\int d(\delta x)\, (\delta x)^2\, P(x)\right]^{3/2}
},
\end{equation}
evaluated over a narrow region surrounding a single bright fringe.
For a symmetric interference pattern, $\mathcal{S}$ vanishes identically.
In the presence of a $\theta$-deformed probability rule, however, the analysis
above implies
\begin{equation}
\mathcal{S}
\propto
\theta\,
\frac{R_1-R_2}{R_1+R_2},
\end{equation}
to leading order, with a proportionality factor set by known geometric and
kinematic parameters.
Importantly, the observable is dimensionless and independent of the absolute
normalization of the probability distribution.

Several features make this skewness a particularly attractive experimental
probe.
First, it is insensitive to global phase drifts and to rigid translations of
the interference pattern, since it is defined locally around each fringe.
Second, symmetric sources of noise---such as finite detector resolution,
phase diffusion, or path-length fluctuations---do not contribute to
$\mathcal{S}$ at leading order.
Third, the signal is enhanced whenever the interfering amplitudes are not
perfectly balanced, a situation that naturally arises in many realistic
interferometric configurations.

From an experimental perspective, the measurement of $\mathcal{S}$ requires
no modification of the underlying interferometric platform.
It amounts to a reanalysis of existing spatially resolved interference data,
with emphasis on the local shape of individual bright fringes rather than on
their global arrangement.
As a result, even null measurements can be used to place direct upper bounds
on the deformation parameter $\theta$, without invoking additional dynamical
assumptions.

More generally, the same logic applies to interferometric protocols in which
interference is encoded in variables other than position, such as time or
frequency.
Whenever a well-defined interference maximum can be identified, its local
asymmetry provides a direct window onto the probabilistic structure of quantum
mechanics.
In this sense, the observable $\mathcal{S}$ offers a minimal and experimentally
transparent route to testing the validity of the Born rule itself.

\section*{Discussion}

The central result of this work is the identification of a local and
symmetry-protected interferometric signature that directly probes the
probabilistic structure of quantum mechanics.
A minimal deformation of the Born rule leaves the linear dynamics,
interference conditions, and fringe positions unchanged, while inducing a
distinct left--right asymmetry in the local shape of bright interference
peaks.
This separation between kinematics and probability allows the effect to be
isolated in a particularly clean and experimentally accessible manner.

From an experimental perspective, the key feature is what does \emph{not}
change.
The deformation does not shift interference maxima, does not modify fringe
spacing, and does not introduce symmetric broadening.
Instead, its leading observable consequence is a cubic, odd-order distortion
of the local intensity profile.
Because conventional sources of noise—such as phase fluctuations,
path-length instability, or finite detector resolution—primarily produce
symmetric effects, the skewness identified here cannot be easily mimicked by
standard experimental imperfections.

The appearance of complex parameters and non-Hermitian operators may invite
comparisons with dissipative or open-system approaches to quantum mechanics~\cite{ElGanainy2018NonHermitian}.
However, the physical interpretation here is fundamentally different.
The present framework does not rely on coupling to an environment, nor does
it invoke effective dynamics obtained by tracing out degrees of freedom.
Instead, the linear Schr\"odinger evolution is preserved exactly, while the
probabilistic assignment itself is modified.
In this sense, the effect identified here should be viewed not as a form of
dissipation, but as a controlled probe of the probability rule underlying
quantum mechanics.

More broadly, the interferometric signature uncovered in this work is not
tied to a specific realization of a $\theta$-deformed Born rule.
Any modification of quantum probability that (i) admits a well-defined
classical limit and (ii) preserves linear superposition at leading order
will necessarily leave the locations and widths of dominant interference
peaks unchanged, while allowing only higher-order, local distortions of
their intensity profiles.
Under these general conditions, deviations from the standard quadratic
probability assignment enter most efficiently through the phase
$\arg\psi$, rather than through the modulus $|\psi|$.

Even in more radical scenarios where linear superposition itself is
ultimately modified, such effects are expected to appear at parametrically
higher order.
From an experimental standpoint, this hierarchy implies that local,
odd-order distortions of interference peaks—such as the skewness identified
here—provide a particularly robust and model-independent probe of the
accuracy of quantum probability.
Precision interferometry can therefore test the probabilistic postulates of
quantum mechanics directly, without requiring detailed assumptions about
the microscopic origin of their possible deformation.

A more systematic theoretical framework underlying the $\theta$-deformed
probability rule has been developed elsewhere~\cite{Zhang:2026gip}.
However, the present analysis does not rely on its details.
The emphasis here is deliberately placed on experimentally accessible
consequences.
In this sense, the observable proposed in this work allows experimental
constraints to be placed on quantum probability independently of how the
underlying theory is ultimately formulated.

Taken together, these considerations suggest a shift in perspective.
Interference experiments do not merely test the superposition principle or
the linear dynamics of quantum mechanics.
They also encode detailed information about the probabilistic interpretation
of quantum amplitudes.
Local asymmetries in interference peaks, long regarded as nonessential or
removable features, may instead provide a sensitive and direct window into
the foundations of quantum probability itself.

\section*{Acknowledgements}
Y.Z. thanks Bin Chen, Song He, Su Yi, and Huaxing Zhu for their interest in this work.
The work of Y.Z. is supported by the National Natural Science Foundation of China under Grant No. 12405086.

\bibliographystyle{physics}

\bibliography{Refs}


\widetext


\onecolumngrid

\begin{center}
{\large\bfseries SUPPLEMENTARY INFORMATION}
\end{center}

\vspace{1.5em}

\twocolumngrid

\appendix

\section*{Derivation of the cubic fringe skewness}

This Supplementary Information provides a detailed derivation of the local cubic skewness discussed in the main text, and is not required for understanding the experimental signatures.

We consider a superposition of two wave packets as defined in
Eq.~\eqref{two_wave_packets}, and work consistently to leading order in the
deformation parameter $\theta$.
The probability density is defined by the $\theta$-deformed Born rule
\eqref{theta_bornrule_arg}, which reduces smoothly to the standard quadratic
Born rule in the limit $\theta\to0$.


Expanding the probability density to first order in $\theta$ yields
\begin{equation}
P = P_0 + \delta P + \mathcal O(\theta^2),
\qquad
\delta P = -2\theta\,P_0\,\arg\psi_0 + \delta P_{\rm lin},
\end{equation}
where $P_0$ denotes the undeformed probability distribution given in \eqref{P0} and
$\arg\psi_0$ is the phase of the superposed wave function evaluated at
$\theta=0$,
\begin{equation}
\arg\psi_0
=
\arg\!\left(
R_1 e^{iS_1/\mathrm{Re}\,\kappa}
+
R_2 e^{iS_2/\mathrm{Re}\,\kappa}
\right).
\end{equation}

The term $\delta P_{\rm lin}$ collects all contributions that are linear in
the classical actions $S_1$ and $S_2$,
\begin{equation}
\delta P_{\mathrm{lin}}
=
\frac{2\theta}{\mathrm{Re}\,\kappa}
\Big[
R_1^2 S_1
+
R_2^2 S_2
+
R_1R_2(S_1+S_2)
\cos\frac{S_1-S_2}{\mathrm{Re}\,\kappa}
\Big].
\end{equation}

Combining the phase-dependent and linear contributions, the full
$\theta$-dependent correction can be written in closed form as
\begin{align}
\delta P
=
\theta \big(R_1^2&-R_2^2\big)
\frac{S_1-S_2}{\mathrm{Re}\,\kappa}
\nonumber
\\ &
-
2 \theta P_0
\arctan \!\left(
\frac{R_1-R_2}{R_1+R_2}
\tan \frac{S_1-S_2}{2 \,\mathrm{Re}\,\kappa}
\right).
\end{align}

For a small amplitude imbalance,
$R_1 = R_0 + \varepsilon$ and $R_2 = R_0 - \varepsilon$ with
$|\varepsilon| \ll R_0$, this expression admits the compact expansion
\begin{equation}
\delta P
=
4 \theta R_0^2 \, \frac{\varepsilon}{R_0}
\Bigg[
\frac{S_1-S_2}{2\,\mathrm{Re}\,\kappa}
-
\sin 
\frac{S_1-S_2}{2\,\mathrm{Re}\,\kappa}
\Bigg]
+
\mathcal O\!\left( \frac{\varepsilon^2}{R_0^2} \right)
+
\mathcal O(\theta^2),
\end{equation}
which is manifestly regular in the symmetric limit and vanishes cubically as
$S_1 \to S_2$.

The analysis above does not rely on the small-asymmetry assumption.
For arbitrary amplitudes $R_1$ and $R_2$, the deformation does not shift the
locations of the interference fringes, which remain fixed by the condition
$
S_1 - S_2 = 2n\pi\,\mathrm{Re}\,\kappa.
$
Instead, the effect of $\theta$ is to modify the local shape of the intensity
profile in the vicinity of each fringe.

To make this explicit, we expand around a bright fringe by writing
\begin{equation}
\frac{S_1 - S_2}{\mathrm{Re}\,\kappa} = 2n\pi + \sigma,
\qquad |\sigma| \ll 1 .
\end{equation}
Near the fringe maximum, the undeformed probability distribution takes the
standard quadratic form
\begin{equation}
P_0
=
(R_1+R_2)^2
-
R_1R_2\,\sigma^2
+
\mathcal O(\sigma^4),
\end{equation}
which is symmetric under $\sigma \to -\sigma$.

The leading $\theta$-dependent correction is found to be
\begin{equation}
\delta P
=
2 n\pi\,\theta\,(R_1^2 - R_2^2)
+
\frac{2}{3}\,
\theta\,R_1R_2
\frac{R_1-R_2}{R_1+R_2}\,
\sigma^3
+
\mathcal O(\sigma^4).
\end{equation}
The first term is a fringe-dependent but phase-independent offset, which may
be absorbed into the overall normalization and does not affect the local
fringe structure.
Discarding this contribution, one arrives at Eq.~\eqref{deltaPsim} in the main
text.

Several features of this result are worth emphasizing.
First, the absence of linear and quadratic terms in $\sigma$ implies that the
fringe position and its curvature are unaffected by the deformation.
Second, the leading correction is cubic and odd under $\sigma\to-\sigma$,
corresponding to a left--right asymmetric distortion of the bright fringe.
Finally, the effect vanishes identically for $R_1 = R_2$, reflecting its
dependence on amplitude imbalance.
These properties follow directly from the fact that the $\theta$ deformation
enters exclusively through the phase $\arg\psi$, and do not depend on the
detailed form of the wave packets.

\end{document}